\documentclass[12pt]{article}
\usepackage{geometry}
\usepackage{graphics}
\usepackage{setspace}
\begin{document}

\title{Perpendicular transport and magnetization processes in magnetic multilayers
with strongly and weakly coupled magnetic layers.}

\author{M.~Zwierzycki\setcounter{footnote}{1}\thanks{corresponding
    author, email: maz@ifmpan.poznan.pl
    } ~and S.~Krompiewski\\
  Institute of Molecular Physics, P.A.S Smoluchowskiego~17, 60-179
  Pozna\'n}

\maketitle
\begin{abstract}
  Within the framework of a two-band tight-binding model, we have
  performed calculations of giant magnetoresistance, exchange coupling
  and thermoelectric power (TEP) for a system consisting of three
  magnetic layers separated by two non-magnetic spacers with the first
  two magnetic layers strongly antiferromagnetically exchange-coupled.
  We have shown how does the GMR relate with the corresponding regions
  of magnetic structure phase diagrams and computed some relevant
  hysteresis loops, too. The GMR may take negative values for specific
  layers thicknesses, and the TEP reveals quite pronounced
  oscillations around a negative bias.
keywords: \emph{}Giant magnetoresistance, Interlayer exchange
coupling, Magnetization processes
\end{abstract}

\section{Introduction}

Since the discovery of giant magnetoresistance (GMR) phenomenon in
magnetic multilayers \cite{Grunberg} there has been a great deal of
interest in studying them by both theoretical and experimental
methods. The reason for the interest is, already partially realized,
possibility of practical applications as magnetic sensors, recording
heads and magnetic memory elements.

The standard system exhibiting the GMR is a trilayer (\emph{i.e.} two
magnetic layers separated by a non-magnetic spacer) with thickness of
the spacer chosen so as to produce antiferromagnetic coupling between
magnetic layers. While such a system, due to its simplicity, is
convenient for theoretical treatment, it presents some problems in
practical applications. The main problem is a high switching field
which is usually necessary to rotate the magnetizations (overcoming
antiferromagnetic coupling) and produce GMR. One way to deal with this
problem is to use somewhat more complex \emph{spin-engineered}
structures. Widely known structures of this type are spin-valve
systems \cite{Dieny}, in which one of the magnetic moments is fixed by
the strong exchange coupling due to an additional antiferromagnetic
layer (\emph{e.g.} MnFe or CoO). There exists however also a different
approach in which a system composed of three magnetic layers is used.
Two of them are strongly antiferromagnetically coupled, forming the so
called artificial antiferromagnetic subsystem (AAF) \cite{Berg}, and
the third one --- detection layer --- is only weakly coupled (or just
decoupled). Such a setup was proposed both for laboratory measurements
\cite{Parkin-Mauri,Bloeman} and more practically as angular velocity
meter \cite{Berg}. A similar system (superlattice with strong and weak
exchange couplings) was also studied theoretically (on \emph{ab
  initio} level) \cite{SKFSUK}, the thicknesses involved were however
small due to numerical limitations.

The aim of the present paper is to perform thorough studies of
transport and magnetic properties of the systems in question and to
relate them with corresponding magnetic structure phase diagrams.

\section{The model and the method of calculations}

We consider a system consisting of three magnetic layers separated by
two non-magnetic spacers, \emph{i.e.} the structure of the \(
F_{1}/S_{1}/F_{2}/S_{2}/F_{3} \) type, where \( F_{i} \) stands for
ferro- and \( S_{i} \) for paramagnetic layer. In order to describe
collinear configurations we employ tight-binding hamiltonian with two,
hybridized bands and spin-dependent on-site potentials (see
Ref.~\cite{SKUK} for details). We restrict ourselves for simplicity to
the case of simple cubic structure. The following values of model
parameters have been chosen: \( E_{F}=0 \), \( t_{s}=-1 \), \(
t_{d}=-0.2 \), \( V_{sd}=1 \), \( \epsilon ^{s}_{i\sigma }=0 \), \(
\epsilon ^{d}_{i\uparrow }=-1 \) for all the layers, \( \epsilon
^{d}_{i\downarrow }=-0.2 \) within magnetic layers and \( -1 \)
elsewhere, where \( E_{F} \) is the Fermi energy, \( t_{\alpha } \)
(with \( \alpha \) being the band index --- \( s,\, d \)) are the
hopping integrals, \( V_{sd} \) is the \( s-d \) intra-atomic
hybridization and \( \epsilon ^{\alpha }_{i\sigma } \) are the on-site
potentials (where \( \sigma \) is \( \uparrow \) for majority- and \(
\downarrow \) for minority-spin carriers). The above set of parameters
enables us to mimic essential features of the electronic structure of
Co/Cu multilayers (\emph{i.e.} the spin-polarized density of states is
qualitatively reproduced --- in particular the majority \emph{d}-bands
in magnetic and \emph{}(non-polarized) \emph{d}-bands in paramagnetic
layers are matched perfectly which closely resembles the situation in
Co/Cu systems).

The conductance is computed from the Kubo formula with the help of
recursion Green function technique \cite{SKUK,Asano}. The only
difference in comparison with \cite{SKUK} is that hybridization in
lead wires (attached to the multilayer from both sides, for transport
calculations) has been taken into account this time. The GMR has been
defined as
\begin{equation}
\label{GMR}
\mathrm{GMR}=\frac{\Gamma ^{\uparrow \downarrow ;\downarrow }}{\Gamma
  ^{\uparrow \downarrow ;\uparrow }}-1,  
\end{equation}
where the arrows show the orientations of magnetic moments. Note that
without the first magnetic layer this definition would be identical
with the usual one.  Additionally the thermopower (TEP) has also been
calculated from the following formula (see eg. Ref.~\cite{Mott})
\begin{equation}
\label{onsanger}
S=-\frac{\pi ^{2}k^{2}_{B}T}{3\left| e\right| }\frac{d}{dE}\,
\mathrm{ln}\, \Gamma \left( E\right) . 
\end{equation}
We define, as in Ref.~\cite{SKUK_thp}, the ``giant
magneto-TEP-effect'' GMTEP analogously to Eq.~(\ref{GMR}) (with \(
\Gamma \) replaced by \( S \)).

For studying the magnetization processes we employ the
phenomenological expression, not unlike the one introduced in
Ref.~\cite{Schmidt}, (\( \Theta _{i} \) is an angle between the
\emph{i}-th magnetic moment and the external field, B)
\begin{eqnarray}
E\left( \Theta _{1},\Theta _{2},\Theta _{3}\right)  & = & -J_{12}\cos
 \left( \Theta _{1}-\Theta _{2}\right) -J_{23}\cos \left( \Theta
 _{2}-\Theta _{3}\right) -J_{13}\cos \left( \Theta _{1}-\Theta
 _{3}\right) \nonumber \\ 
 &  & -B\sum ^{3}_{i=1}t_{i}\cos \left( \Theta _{i}\right) /t+\sum
 ^{3}_{i=1}E_{A}(\Theta _{i}),\label{phen_energy}  
\end{eqnarray}
where the first three terms describe bilinear exchange coupling
between magnetic layers, the next three are Zeeman energy terms (\(
t_{i} \) being \( i \)-th layer thickness and \( t \) the overall
thickness of all the magnetic layers) and \( E_{A}(\Theta _{i}) \) is
the crystalline anisotropy, that is \( t_{i}K\sin ^{2}(2\Theta
_{i})/4t \) and \( -t_{i}D\cos ^{2}\Theta _{i}/t \) for cubic and
uniaxial case respectively.  We assume that external magnetic field is
applied along the {[}10{]} in-plane crystallographic axis, which can
be either easy or hard axis depending on the sign of the anisotropy
constants. The magnetic moments are confined to the layers plane which
corresponds to the strong shape anisotropy. Expression
(\ref{phen_energy}) was then numerically minimized, with respect to
the \( \Theta _{i} \)-s, by taking, starting from initial
configuration, little steps in the direction opposite to the energy
gradient. All the extremal points found in this way were additionally
checked against the stability condition (\emph{i.e.} the positivity of
all the minors of \( M_{ij}=\partial ^{2}E/(\partial \Theta
_{i}\partial \Theta _{j}) \)) in order to eliminate saddle points.
Using Eq.~(\ref{phen_energy}) one can write (for \( B \) equal to 0)
\begin{eqnarray}
J_{12} & = & \frac{1}{4}\left[ E\left( \pi 00\right) +E\left( 0\pi
    0\right) -E\left( 000\right) -E\left( 00\pi \right) \right]
,\nonumber \\ 
J_{23} & = & \frac{1}{4}\left[ E(00\pi )+E\left( 0\pi 0\right)
    -E\left( 000\right) -E\left( \pi 00\right) \right] ,\label{joty}
    \\ 
J_{13} & = & \frac{1}{4}\left[ E\left( 00\pi \right) +E\left( \pi
    00\right) -E\left( 000\right) -E\left( 0\pi 0\right) \right]
    .\nonumber  
\end{eqnarray}
Therefore, having known the energies of collinear configurations from
the model calculations (based on the two-band tight-binding
hamiltonian), we are able to determine the exchange coupling
constants.

From now on we will be using reduced values of the magnetic field \(
b=B/|J_{12}| \) and the anisotropy constants \( k=K/|J_{12}| \) and \(
d=D/|J_{12}| \). We will also assume, if not stated explicitly
otherwise, the magnetic layers thicknesses to be 8, 3 and 3~ML
(monolayers), respectively, in order to keep the length ratios as in
Ref.~\cite{Bloeman}. The first spacer thickness will be set to 3~ML in
order to achieve the needed antiferromagnetic coupling between the
first two magnetic layers.

\section{Results}

\begin{figure}
{\par\centering \resizebox*{!}{11cm}{\includegraphics{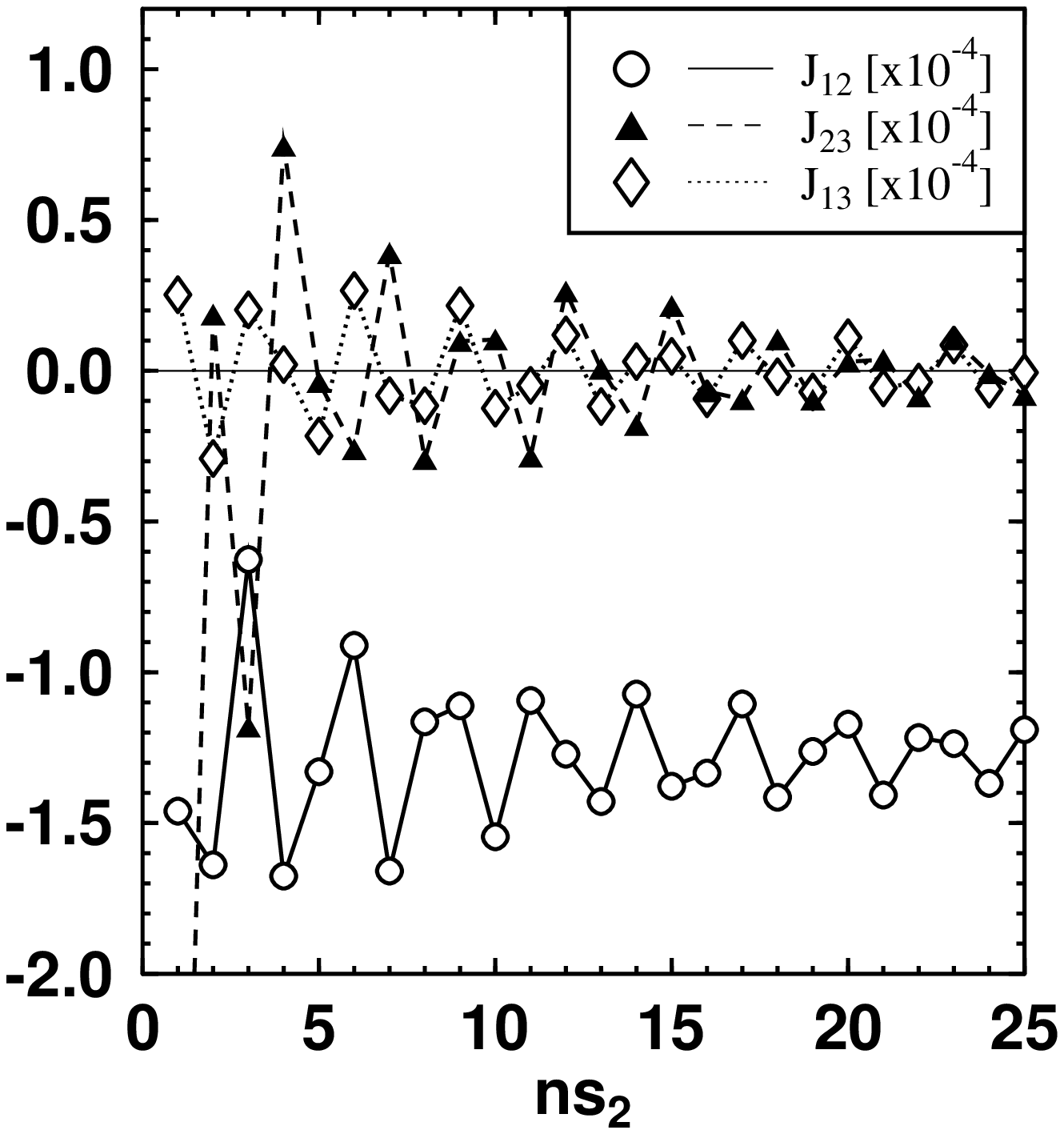}} \par}

\caption{Exchange coupling constants, calculated for the \protect\( 
  F_{1}/S_{1}/F_{2}/S_{2}/F_{3}\protect \) structure, plotted against
  the second spacer thickness (\protect\( ns_{2}\protect \)).
  Magnetic layer thicknesses have been set to 8, 3 and 3~ML,
  respectively and the first spacer thickness is equal to 3~ML. The
  hamiltonian parameters are as follows: \protect\( E_{F}=0\protect \)
  (Fermi energy), \protect\( t_{s}=-1\protect \), \emph{\protect\(
    t_{d}=-0.2\protect \)} (hopping integrals), \protect\(
  V_{sd}=1\protect \) (hybridization), \protect\( \epsilon _{i\sigma
    }^{s}=0\protect \) and \protect\( \epsilon _{i\uparrow
    }^{d}=-1\protect \)(for all the layers), \protect\( \epsilon
  _{i\downarrow }^{d}=-0.2\protect \) for \protect\( i\protect \)
  within the magnetic layer and \protect\( -1\protect \) elsewhere.
  Note the strong antiferromagnetic coupling (\protect\(
  J_{12}\protect \)) between the first two magnetic
  layers.\label{exch_pl}}
\end{figure}

\begin{figure}
{\par\centering \resizebox*{!}{11cm}{\includegraphics{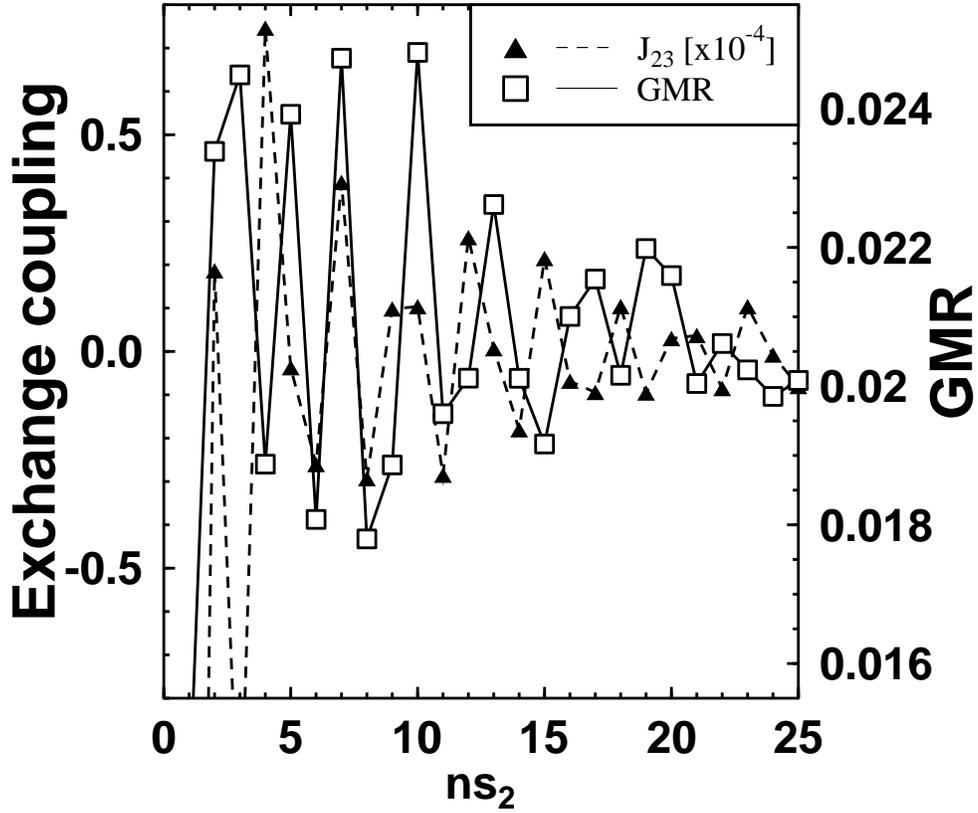}} \par}

\caption{CPP-GMR and the interlayer exchange coupling 
  (\protect\( J_{23}\protect \)) plotted as a function of the second spacer
  thickness. The period of oscillation is about 3~ML in agreement with
  the predicted value of 2.8~ML, resulting from the corresponding
  Fermi surface calliper.  Asymptotically the oscillations of both
  quantities are predominantly opposite in phase.\label{gmr_pl}}
\end{figure}

\begin{figure}
{\par\centering \resizebox*{!}{9cm}{\includegraphics{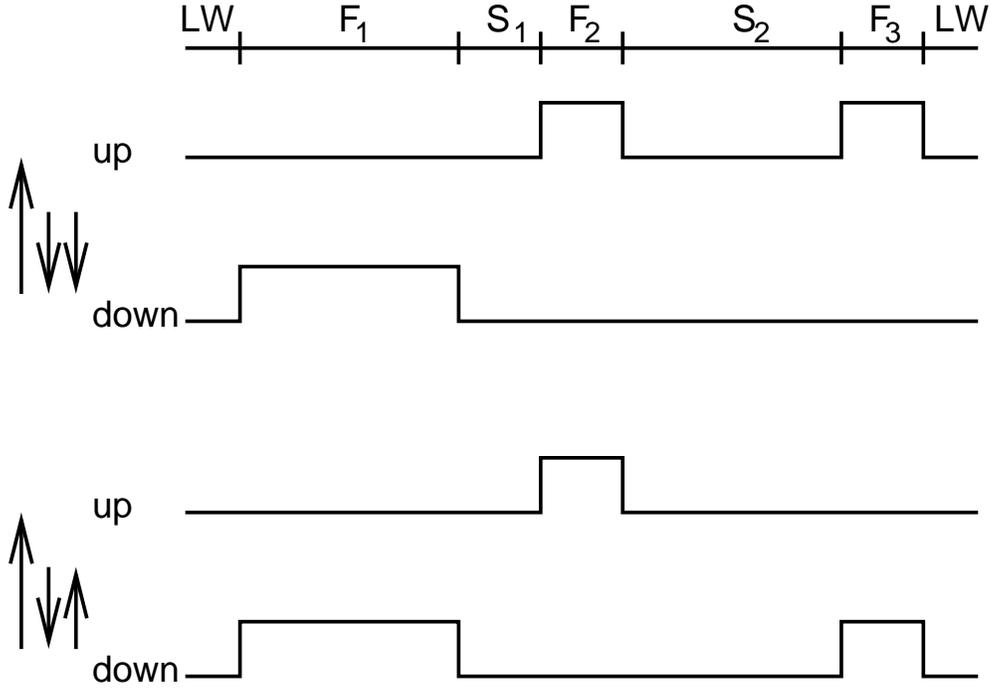}} \par}

\caption{Schematic plot of the on-site potentials for \emph{d}-bands
  (\protect\( \epsilon _{i\sigma }^{d}\protect \)) across the
  \protect\( F_{1}/S_{1}/F_{2}/S_{2}/F_{3}\protect \) multilayer for
  different configurations and spin directions. The \protect\(
  \uparrow \downarrow ;\downarrow \protect \) down- and \protect\(
  \uparrow \downarrow ;\uparrow \protect \) up-spin channels with only
  two scattering interfaces are clear candidates for high conductance
  channels. Note that there is however no non-scattering channel for
  configurations taken into account.\label{d-band_pl}}
\end{figure}

\begin{figure}
{\par\centering \resizebox*{!}{11cm}{\includegraphics{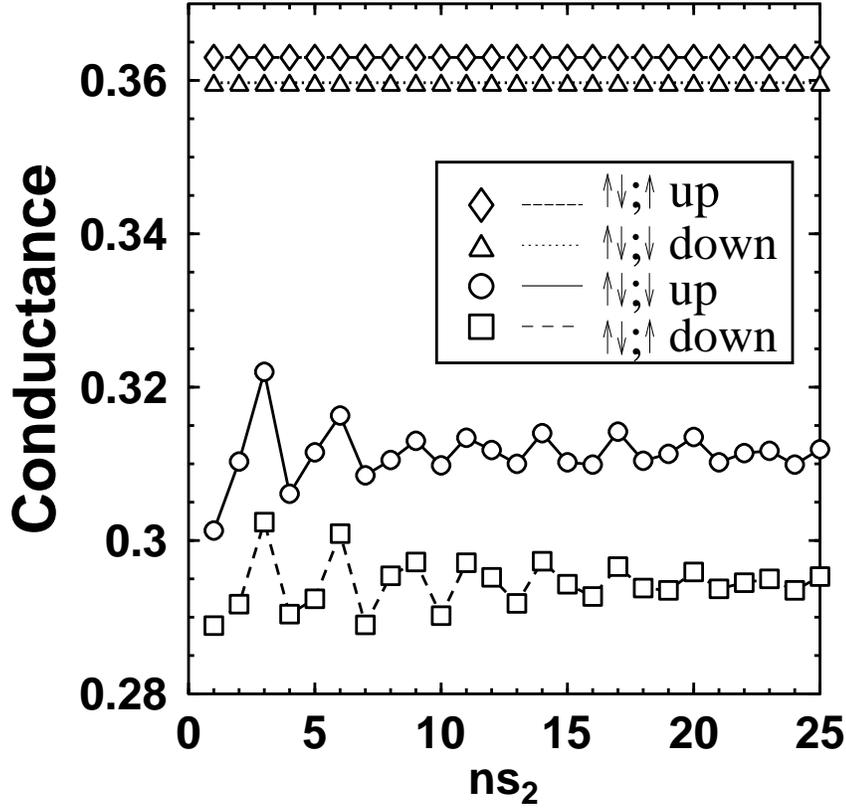}} \par}

\caption{Conductances (in \protect\( e^{2}/h\protect \) units) for
  parameters as in Fig.~\ref{gmr_pl}. There exist, as discussed in the
  text, two high and two low conductance channels.\label{gamma_pl}}
\end{figure}

\begin{figure}
{\par\centering \resizebox*{!}{11cm}{\includegraphics{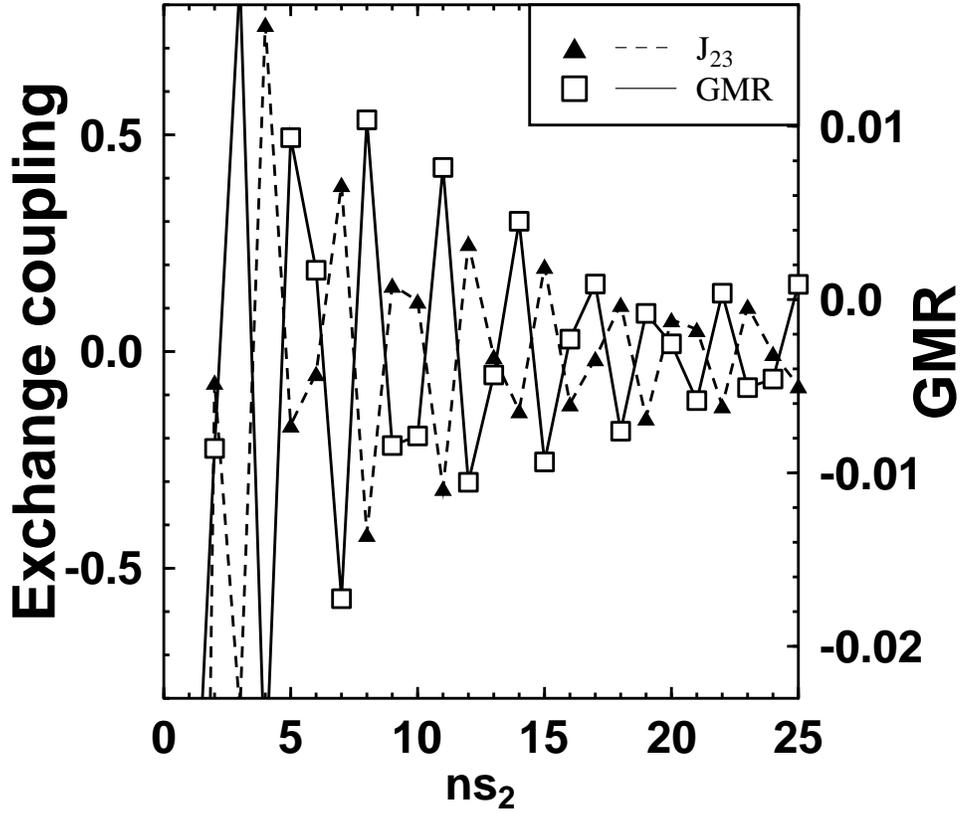}} \par}

\caption{CPP-GMR and the interlayer exchange coupling 
  (\protect\( J_{23}\protect \)) calculated for the same set of
  parameters as in Fig.~\ref{gmr_pl}, except for the thickness of the
  second magnetic layer (\protect\( nf_{2}\protect \)) which has been
  set to 5~ML. The GMR oscillate around a small \emph{negative}
  value.\label{gmr1_pl}}
\end{figure}

Figure~\ref{exch_pl} presents the exchange coupling constants (\( J
\)-s) plotted against the thickness of the second spacer (\( ns_{2}
\)). As mentioned above, for the chosen thicknesses we got strong
antiferromagnetic coupling (\( J_{12} \)) between the first two
magnetic layers, while \( J_{23} \) and \( J_{13} \) oscillate around
zero. In all three cases the period of oscillations is about 3~ML,
which is close to the theoretically predicted value (2.8~ML) coming
from the stationary spanning vector \cite{Bruno,SKMZUK} placed at the
\( (0,\pi ) \) (and equivalent positions) in the two-dimensional
Brillouin zone. The second period (8~ML) originating from the hole
pocket placed at \( (\pi ,\pi ) \) does not seem to appear in the
present context. Note however that it can become visible under some
circumstances like in Ref.~\cite{SKUK} where the tunnelling
conductance has been considered. The GMR and \( J_{23} \) for the same
system have been plotted in Fig.~\ref{gmr_pl}. It can be noted that
GMR asymptotically tends to oscillate with the same period but in
opposite phase to \( J_{23} \).  This is in agreement with our
previous findings \cite{SKMZUK}, but it is still not clear to what
extent this correlation is universal. The values of GMR are strongly
reduced in comparison with the trilayer case. This can be easily
understood if we note that, due to the fixed antiferromagnetic
alignment of the first two magnetic layers, there is no non-scattering
channel in any of the configurations involved (see Eq.~\ref{GMR}).
Figure~\ref{d-band_pl} where the on-site potentials for the
\emph{d}-bands (\( \epsilon ^{d}_{i\sigma } \)) have been
schematically plotted, shows that there exist at least two scattering
interfaces in each case.  Basing on the number of interfaces one can
qualitatively predict that the \( \uparrow \downarrow ;\downarrow \)
down- and \( \uparrow \downarrow ;\uparrow \) up-spin electron
channels have the higher conductances than the remaining two. This is
indeed clearly visible in Fig.~\ref{gamma_pl} where the computed
conductances are shown. As already discussed there is no obvious
highest conductance channel. Instead, we have two higher and two lower
conducting channels close to each other within the pairs. As a
consequence the sign of GMR is determined by all the channels, and can
be changed by manipulating some parameters (\emph{eg.} thicknesses of
the layers). This is the case in Fig.~\ref{gmr1_pl} where we have
plotted the GMR and \( J_{23} \) for a system with thickness of the
second magnetic layer set to 5~ML (instead of 3~ML as in
Fig.~\ref{gmr_pl}). The GMR oscillations have a small but clearly
negative bias. The asymptotic opposite-phase correlation with respect
to \( J_{23} \) is again clear in this case.

\begin{figure}
{\par\centering \resizebox*{!}{11cm}{\includegraphics{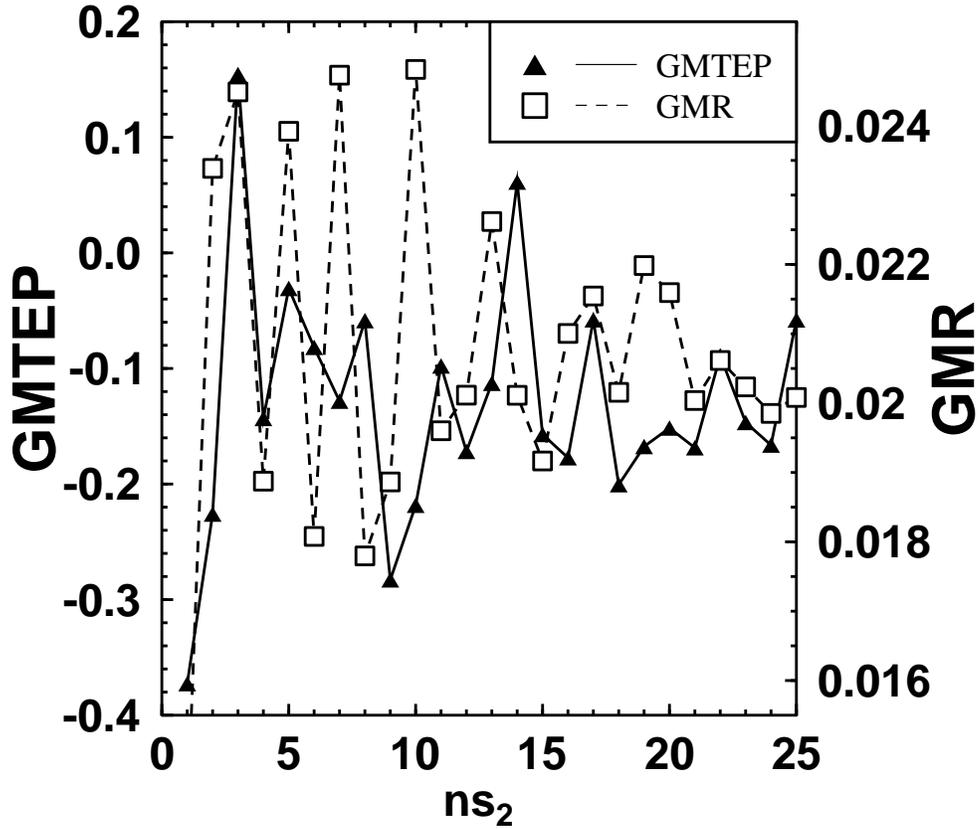}} \par}

\caption{GMTEP (giant magneto-TEP effect) for the same set of
  parameters as in Fig.~\ref{gmr_pl}.  GMR has been also plotted for
  reference. The GMTEP oscillates asymptotically in roughly the same
  phase but around a negative bias.\label{gmtep_pl}}
\end{figure}

The GMTEP, calculated for the same set of parameters as in
Fig.~\ref{gmr_pl}, has been plotted in Fig.~\ref{gmtep_pl}. In
agreement with the findings of Ref.~\cite{SKUK_thp}, the oscillations
are quite pronounced and have the same period as GMR but exhibit
negative bias. Asymptotically they seem to have roughly the same phase
as GMR.

For studying the magnetization processes we have chosen the parameters
as in Fig.~\ref{exch_pl} with the second spacer thickness (\( ns_{2}
\)) set to 7~ML (the second ferromagnetic maximum of \( J_{23} \)). As
already stated, two cases have been taken into account, \emph{i.e.}
the cubic and uniaxial anisotropies.

\begin{figure}
{\par\centering \resizebox*{!}{6cm}{\includegraphics{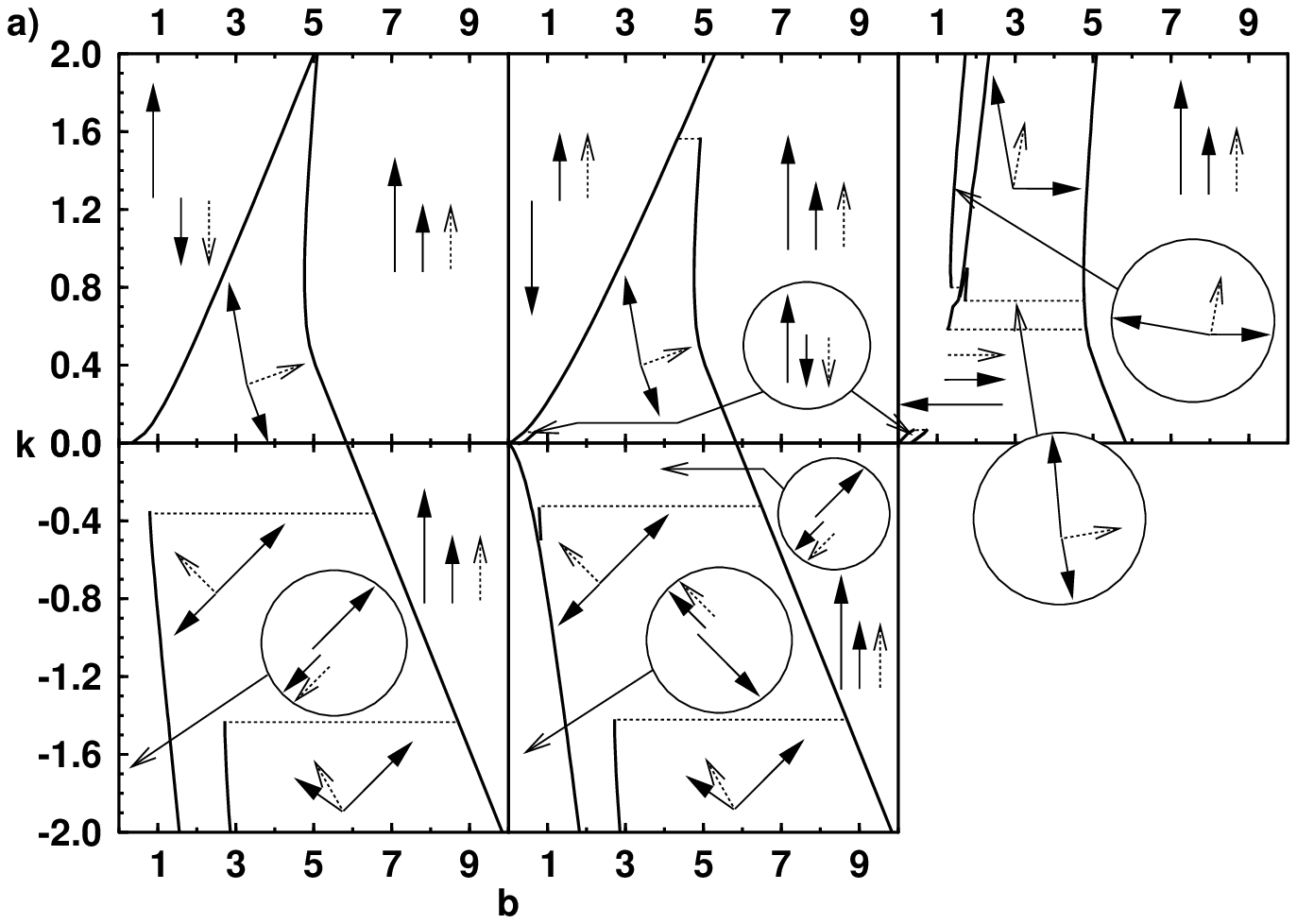}}  \resizebox*{!}{6cm}{\includegraphics{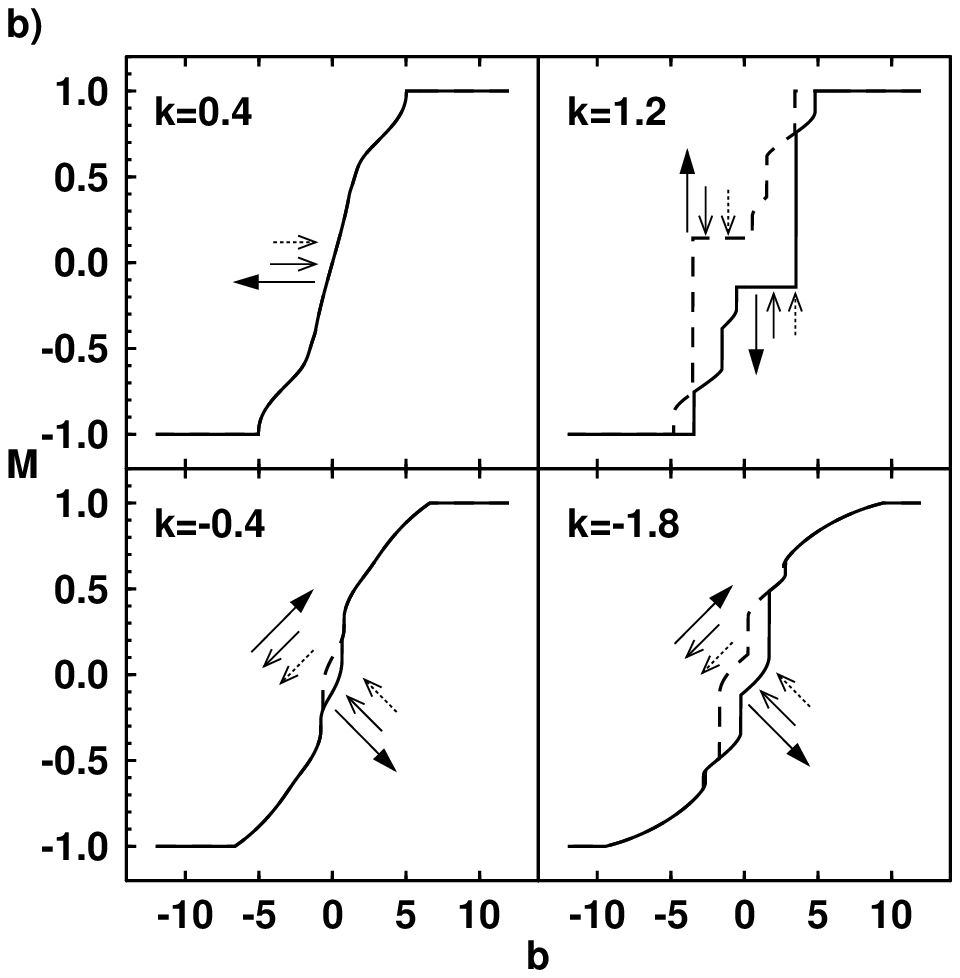}} \par}

\caption{a) Phase diagrams for magnetization processes obtained for
  different initial configurations in the case of cubic anisotropy.
  The second spacer thickness (\protect\( ns_{2}\protect \)) has been
  set to 7~ML. The arrangements of magnetic moments are schematically
  drawn with the first (8~ML) magnetic layer symbolized by the long
  solid arrow and the third one (detection layer) by the dotted arrow.
  b)~Exemplary hysteresis loops with the dashed line marking the
  backward sweep.  The zero field configurations are schematically
  marked for both the sweep directions.  For positive values of
  \protect\( k\protect \), the strong anisotropy produces flat regions
  typical for exchange-biased systems.\label{phase-cubic_pl}}
\end{figure}

In the first case the anisotropy term (\( t_{i}k\sin ^{2}(2\Theta
_{i})/4t \)) gives rise to four potential wells placed at the
following in-plane crystallographic axes~: {[}10{]}, {[}01{]},
{[}\=10{]} and {[}0\=1{]} for \( k>0 \) and {[}11{]}, {[}1\=1{]},
{[}\=1\=1{]} and {[}\=11{]} for \( k<0. \)
Figure~\ref{phase-cubic_pl}a exemplifies some phase diagrams for
various initial configurations. We have chosen the simplest ones
\emph{i.e.} the collinear configurations with relative alignment of
magnetic moments favoured by interlayer exchange coupling. It is
however, in principal, possible to stabilize also non-collinear ones,
provided that the anisotropy is strong enough. The phase diagrams have
been obtained by taking subsequent scans along \( b \) for different
\( k \) values. Dotted horizontal lines are thus only guides to the
eye. The diagrams exhibit rich structure with a number of
configurations (phases) occurring during the magnetization process
(including the non-collinear configurations). The transition between
them can be either of first or second type, that is they manifest
themselves as discontinuities in magnetization or its first
derivative. Some exemplary hysteresis loops are presented in
Fig.~\ref{phase-cubic_pl}b. As expected stronger anisotropy produces a
richer structure. For positive, and sufficiently big, values of \( k
\) some flat regions, typical for exchange-biased systems, occur.
Note however that there is no \( \uparrow \downarrow ;\downarrow \) to
\( \uparrow \downarrow ;\uparrow \) transition in the first diagram of
Fig.~\ref{phase-cubic_pl}a (but see below).

\begin{figure}
{\par\centering \resizebox*{!}{3.5cm}{\includegraphics{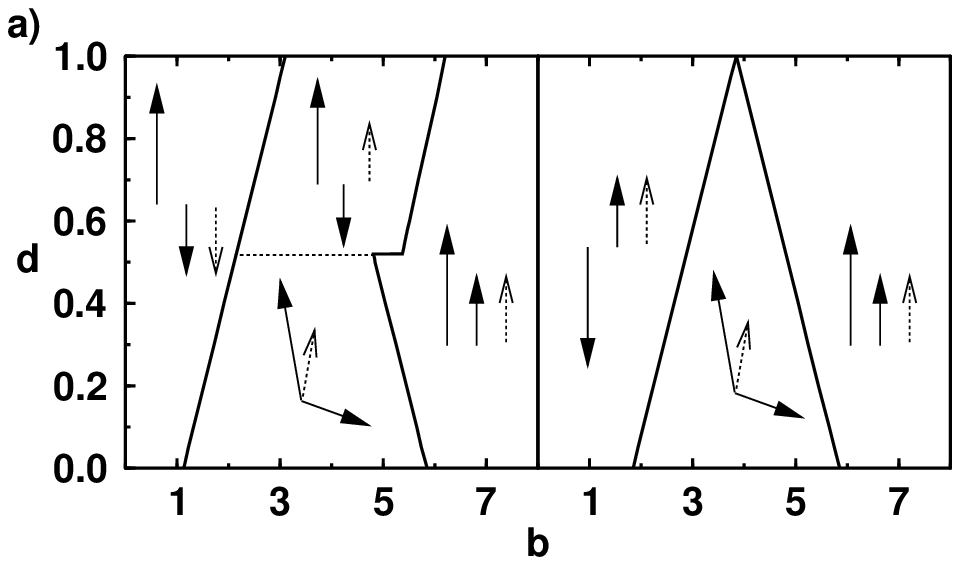}} 
\resizebox*{!}{3.5cm}{\includegraphics{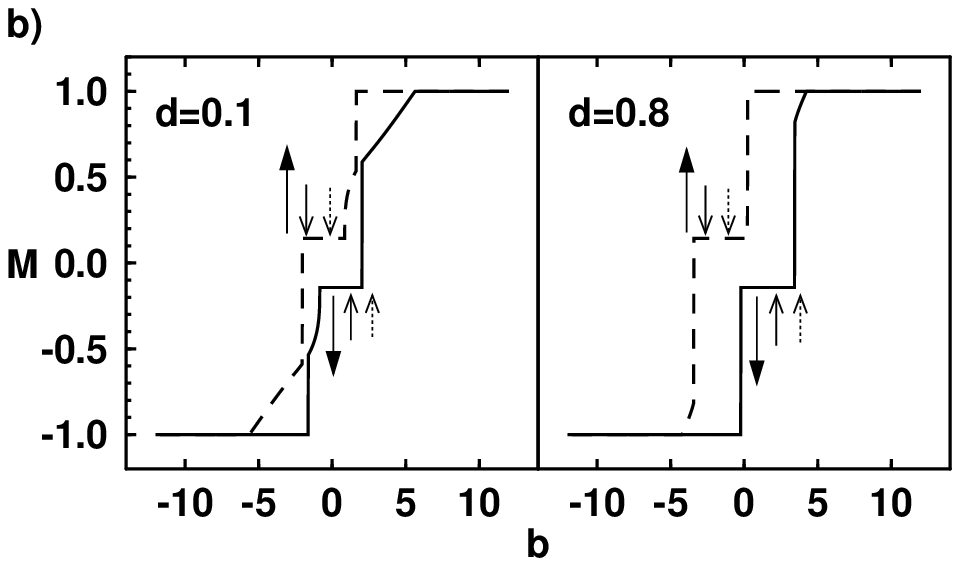}} \par}

\caption{a) Phase diagrams in the uniaxial anisotropy
  case. b)~Exemplary hysteresis loops. Only positive values of
  \emph{d} are shown since the opposite case exhibits no hysteresis at
  all. The flat regions are present already for weak
  anisotropy.\label{phase-uni_pl}}
\end{figure}

For the case of uniaxial anisotropy (\( -t_{i}d\cos ^{2}(\Theta
_{i})/t \)) there exist only two potential wells, \emph{}that is
{[}10{]} and {[}\=10{]} for \( d>0 \) and {[}0\=1{]}, {[}01{]} for \(
d<0 \). The phase diagrams (Fig.~\ref{phase-uni_pl}a) are somewhat
less complicated now, due to the simpler energy landscape. The
saturation curve for the upper part of the first diagram (with the \(
\uparrow \downarrow ;\downarrow \) initial configuration) has been
obtained on analytical basis (\emph{i.e.} from the stability
condition) because of the weak stability of the \( \uparrow \downarrow
;\uparrow \) configuration (by which we mean that the existing energy
minimum is very shallow) which makes it difficult to perform reliable
numerical minimalization. The same precautions have been applied to
the first hysteresis loop in Fig.~\ref{phase-uni_pl}b.  Note that this
time the flat regions are present already for small values of \( d \).
For \( d>0.51 \) there exists, in the first diagram, the above
mentioned \( \uparrow \downarrow ;\downarrow \) to \( \uparrow
\downarrow ;\uparrow \) transition. Only positive values of \( d \)
were presented since the opposite case is trivial --- there is
practically no hysteresis.

\section{Conclusions}

Within the microscopic two-band tight-binding model we have performed
the calculations of interlayer exchange coupling,
current-perpendicular-to-plane conductance and thermopower for a
system consisting of three magnetic layers, separated by paramagnetic
ones. With the thicknesses chosen so as to produce strong
antiferromagnetic coupling between the first two magnetic layers, we
found both the interlayer exchange coupling, GMR and GMTEP to
oscillate, as a function of the second spacer thickness, with the
period originating from one of the extremal spanning vectors of the
Fermi surface. Additionally, using the phenomenological approach, we
have computed magnetic phase diagrams and commented on their relevance
to the magnetoresistance. We found that the phase diagrams exhibit
rich structure and there are flat regions in hysteresis loops, typical
for exchange-biased spin-valves.  It has been also found that in the
case of the systems under consideration, in contrast to conventional
trilayers, it is possible to obtain a negative (inverse) perpendicular
GMR by merely changing thicknesses of particular layers.

\section{Acknowledgments}

The KBN grants 2PO3B-118-14 and 2PO3B-117-14 are gratefully
acknowledged. We also thank Pozna\'n Computing and Networking Center
for the computing time.

\listoffigures

\end{document}